\documentclass[twocolumn,showpacs,superscriptaddress,amsmath,amssymb,prl]{revtex4-1}
\usepackage[colorlinks,linkcolor=blue,anchorcolor=blue,citecolor=blue,urlcolor=blue]{hyperref}
\usepackage{graphicx}
\usepackage{epsfig}
\usepackage{dcolumn}
\usepackage{bm}
\usepackage{overpic}
\usepackage{color}
\usepackage{multirow}
\usepackage{subfigure}
\usepackage{lineno}
\usepackage{verbatim}
\usepackage{cleveref}
\usepackage{amsthm,amsmath,amssymb}
\usepackage{mathrsfs}
\usepackage{booktabs}

\newcommand{\psip}{\psi(3686)}

\newcommand{\OOb}{\Omega^-\bar{\Omega}^{+}}

\newcommand{\BESIIIorcid}[1]{\href{https://orcid.org/#1}{\hspace*{0.1em}\raisebox{-0.45ex}{\includegraphics[width=1em]{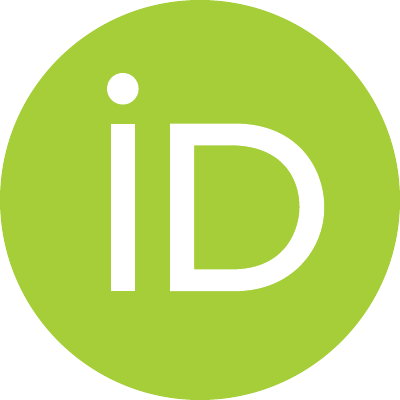}}}} 

\begin{document}
\hyphenpenalty=10000
\tolerance=1000

\title{Measurement of CP Asymmetry Parameters and Polarization Correlations in $\Omega^{-}\bar{\Omega}^{+}$ Pairs}

\author{
\begin{center}
M.~Ablikim$^{1}$\BESIIIorcid{0000-0002-3935-619X},
M.~N.~Achasov$^{4,c}$\BESIIIorcid{0000-0002-9400-8622},
P.~Adlarson$^{84}$\BESIIIorcid{0000-0001-6280-3851},
X.~C.~Ai$^{90}$\BESIIIorcid{0000-0003-3856-2415},
C.~S.~Akondi$^{32A,32B}$\BESIIIorcid{0000-0001-6303-5217},
R.~Aliberti$^{40}$\BESIIIorcid{0000-0003-3500-4012},
A.~Amoroso$^{83A,83C}$\BESIIIorcid{0000-0002-3095-8610},
Q.~An$^{79,66,\dagger}$,
Y.~H.~An$^{90}$\BESIIIorcid{0009-0008-3419-0849},
M.~S.~Anderson$^{40}$\BESIIIorcid{0009-0008-1550-2632}
% ...
\\
\vspace{0.2cm}
(BESIII Collaboration)\\
\vspace{0.2cm} {\it
$^{1}$ Institute of High Energy Physics, Beijing 100049, People's Republic of China\\
% ...
$^{4}$ Budker Institute of Nuclear Physics SB RAS (BINP), Novosibirsk 630090, Russia\\
% ...
$^{32}$ INFN Sezione di Ferrara, (A)INFN Sezione di Ferrara, I-44122, Ferrara, Italy; (B)University of Ferrara, I-44122, Ferrara, Italy\\
% ...
$^{40}$ Johannes Gutenberg University of Mainz, Johann-Joachim-Becher-Weg 45, D-55099 Mainz, Germany\\
% ...
$^{66}$ State Key Laboratory of Particle Detection and Electronics, Beijing 100049, Hefei 230026, People's Republic of China\\
% ...
$^{79}$ University of Science and Technology of China, Hefei 230026, People's Republic of China\\
% ...
$^{83}$ University of Turin and INFN, (A)University of Turin, I-10125, Turin, Italy; (B)University of Eastern Piedmont, I-15121, Alessandria, Italy; (C)INFN, I-10125, Turin, Italy\\
$^{84}$ Uppsala University, Box 516, SE-75120 Uppsala, Sweden\\
% ...
$^{90}$ Zhengzhou University, Zhengzhou 450001, People's Republic of China\\

\vspace{0.2cm}
$^{\dagger}$ Deceased\\
% ...
$^{c}$ Also at the Novosibirsk State University, Novosibirsk, 630090, Russia\\
% ...
}

\end{center}
\vspace{0.4cm}
}

\begin{abstract}

Using $(2.71 \pm 0.01) \times 10^9$ $\psi(3686)$ events collected with the BESIII detector, a joint full angular distribution analysis is carried out for the process $\psi(3686) \to \Omega^-(\to\Lambda K^-) \, \bar{\Omega}^{+}(\to \bar{\Lambda}K^+)$. The first simultaneous measurement of the weak decay parameters $\phi_{\Omega^{-}}$ and $\phi_{\bar{\Omega}^{+}}$ for $\Omega^- \to K^-\Lambda$ and $\bar{\Omega}^+ \to K^+\bar{\Lambda}$ is performed, yielding the first result for the CP-sensitive observable, $\phi_{\rm CP} = (-0.004 \pm 0.055 \pm 0.017)~\text{rad}$, where the first and second uncertainties are statistical and systematic, respectively.
This further enables the extraction of the weak and strong phase differences between the $P$- and $D$-wave amplitudes: $(\xi_D - \xi_P) = (-0.15 \pm 2.25 \pm 0.69)~\text{rad}$ and $(\delta_D - \delta_P) = (-0.97 \pm 0.88 \pm 0.34)~\text{rad}$. Additionally, the polarization correlations between $\Omega^{-}$ and $\bar{\Omega}^{+}$ are measured.

 

\end{abstract}

\maketitle

The Standard Model, despite its remarkable success, fails to explain the fundamental matter-antimatter asymmetry observed in the universe~\cite{Bernreuther:2002uj,Canetti:2012zc}. This is one of cosmology’s greatest unsolved puzzles, and as elucidated by A.D. Sakharov, it requires three essential conditions  for dynamically generating the asymmetry~\cite{Sakharov:1967dj}. Among them is CP violation (CPV), the violation of the combined charge-conjugation and parity symmetry.~\cite{Rubakov:1996vz,Morrissey:2012db}.
Although the Standard Model predicts small CPV~\cite{Cabibbo:1963yz,Kobayashi:1973fv} and it has been well established in meson decays~\cite{Christenson:1964fg,Belle:2001zzw,LHCb:2019hro}, the effects observed to date, including the first observation in baryon decays by LHCb with 5.2$\sigma$ significance~\cite{LHCb:2025ray}, are insufficient to account for the matter-antimatter asymmetry of the universe~\cite{Sakharov:1967dj,Morrissey:2012db,Gavela:1994dt}. Therefore, searches for new sources of CPV remain important. Investigating CPV in higher-spin, multi-strange baryons such as the $\Omega^{-}$ baryon, provides a unique perspective for CPV test since it is the only observed spin-3/2 baryon that decays purely via the weak interaction. Nevertheless, experimental studies of the $\Omega$ baryon remain scarce.

In the weak hyperon decays, 
CP symmetry is tested by comparing the angular distributions of a decay and its charge-conjugate process, with CPV asymmetries arising from interference among the contributing decay amplitudes. In the $\Omega^{-}\to K^{-}\Lambda$ process, the amplitude is described by a parity-conserving ($P$-wave) and a parity-violating ($D$-wave)
amplitude, quantified in terms of the decay parameters $\alpha$, $\beta$, and $\gamma$~\cite{Bacchetta:2000jk,Commins:1998jv,Lee:1957qs}. These parameters are constrained by the normalization condition:
\begin{equation}\label{normalization}
\begin{aligned}
 \alpha^{2} + \beta^{2} + \gamma^{2} = 1,
\end{aligned}
\end{equation} 
where $\beta$ and $\gamma$ are expressible in terms of $\alpha$ and the relative phase $\phi$ between the parity-conserving and parity-violating amplitudes: 
\begin{equation}\label{decay_parameter}
\begin{aligned}
  \beta= \sqrt{1-\alpha^2}\sin\phi, \quad 
  \gamma = \sqrt{1-\alpha^2}\cos\phi,\\
\end{aligned}
\end{equation} 
while the antiparticle has the corresponding decay parameters $\bar{\alpha}$ and $\bar{\phi}$. To construct tests of CP symmetries, we use the following CP-violation observables~\cite{Donoghue:1986hh,Donoghue:1985ww} to compare the particle and antiparticle decay parameters:
\begin{equation}\label{ACP}
A_{\rm CP} = \frac{\alpha+\bar{\alpha}}{\alpha-\bar{\alpha}}, \quad
B_{\rm CP} = \frac{\beta+\bar{\beta}}{\alpha-\bar{\alpha}}, \quad
\phi_{\rm CP} = \frac{\phi+\bar{\phi}}{2}.
\end{equation}

The $\Omega^{-}$ baryon predominantly decays through the $\Omega^{-}\to K^{-}\Lambda$ and $\Omega^{-}\to \Xi\pi$ channels ($\Xi^{0}\pi^{-}$ and $\Xi^{-}\pi^{0}$)\cite{BESIII:2023ldd}, all of which receive contributions from both $P$-wave and $D$-wave amplitudes. A non-vanishing direct CP-violating asymmetry in these decays requires two essential conditions: (1) the interference between different partial waves (e.g., $P$-$D$ wave interference within a given decay channel); and (2) the coexistence of both CP-even and CP-odd amplitudes that acquire non-zero relative phase differences. This mechanism is analogous to that established in $K\to\pi\pi$ decays~\cite{Christenson:1964fg}, where CPV requires both a weak phase (from the Cabibbo-Kobayashi-Maskawa matrix) and a strong phase (from final-state interactions)~\cite{,Kobayashi:1973fv,Watson:1954uc}. Therefore, the CP asymmetries $A_{\rm CP}$ and $B_{\rm CP}$ must incorporate both the strong-phase difference ($\delta_{D}-\delta_{P}$) and the weak-phase difference ($\xi_{D}-\xi_{P}$) between partial waves in each decay channel~\cite{Tandean:2004mv,Salone:2022lpt,Tandean:2002vy}. Actually, a full description of the decay amplitudes would require the inclusion of channel-coupling effects~\cite{Tandean:2004mv}. According to the theoretical model in Ref.~\cite{Tandean:2004mv}, the CP-violating asymmetry in the $\Omega^{-}\to\Xi\pi$ decay channel is predicted to be very small. This is primarily because the weak-phase difference in the $\Omega^{-}\to\Xi\pi$ channel is calculated to be much smaller than in the $\Omega^{-}\to\Lambda K^{-}$ channel, and its strong-phase difference also leads to a suppressed CP-violating effect. Consequently, the $\Omega^{-}\to\Xi\pi$ channel has a negligible impact on the extraction of ($\xi_{D}-\xi_{P}$) and ($\delta_{D}-\delta_{P}$) at leading order. This justifies an approach where these parameters are estimated using only the decay parameters from $\Omega^{-}\to K^{-}\Lambda$ decays~\cite{Tandean:2004mv}:
\begin{equation}\label{weakCP}
\begin{aligned}
 \tan(\delta_{D}-\delta_{P}) \approx\frac{\beta-\bar{\beta}}{\alpha-\bar{\alpha}},\quad 
 \tan(\xi_{D}-\xi_{P}) \approx \frac{\beta+\bar{\beta}}{\alpha-\bar{\alpha}}.
\end{aligned}
\end{equation}

All CP asymmetries defined above depend on the parameters $\alpha_{\Omega^{-}}$, $\alpha_{\bar{\Omega}^{+}}$, $\phi_{\Omega^{-}}$, and $\phi_{\bar{\Omega}^{+}}$. Previous measurements determined $\alpha_{\Omega^{-}} = 0.0154 \pm 0.002$ and $\alpha_{\bar{\Omega}^{+}} = -0.018 \pm 0.004$ with high precision~\cite{Chen:2005aza,Lu:2005fc,Lu:2006bn}. The angle $\phi_{\Omega}$ remained unmeasured until the BESIII study~\cite{BESIII:2020lkm}, although that study did not account for CP violation in $\Omega^-\to K^-\Lambda$.

Beyond CP asymmetries, the spin-entangled production of $\Omega^-\bar{\Omega}^+$ pairs in $e^{+}e^{-}\to\psip\to\Omega^{-}\bar{\Omega}^{+}$ provides a unique opportunity to study polarization correlations between a spin-3/2 baryon and its antibaryon. The polarization of a single $\Omega^-$ has been studied via the decay chain $\Omega^{-}\to K^{-}\Lambda, \Lambda\to p\pi^{-}$~\cite{Zhang:2023box,Zhang:2023wmd,Perotti:2018wxm}. Unlike spin-1/2 particles with a spin vector and two form factors~\cite{BESIII:2018cnd,BESIII:2022qax,BESIII:2020fqg,BESIII:2021ypr,BESIII:2023drj,BESIII:2023lkg}, spin-3/2 systems provide richer polarization information, including a quadrupole and an octupole tensor (fifteen independent components) parameterized by three helicity amplitude ratios $h_i$ and $\phi_i$ ($i=1,3,4$)~\cite{Zhang:2023box,Zhang:2023wmd,Doncel:1972ez,Dubnickova:1992ii}. However, the correlations between $\Omega^-$ and $\bar{\Omega}^+$ in such an entangled state have never been measured. Simultaneous reconstruction of the pair should reveal significant polarization correlations between the two particles~\cite{Zhang:2023wmd}.
To address this, we adopt the approach previously employed in other hyperon decay analyses~\cite{BESIII:2026ala}, but now applied to the CP estimation in the process  $e^{+}e^{-}\to\psip\to\Omega^{-}\bar{\Omega}^{+}$, allowing for a simultaneous description of all decay sequences.

\begin{figure}
 \centering
   \includegraphics[width=0.45\textwidth]{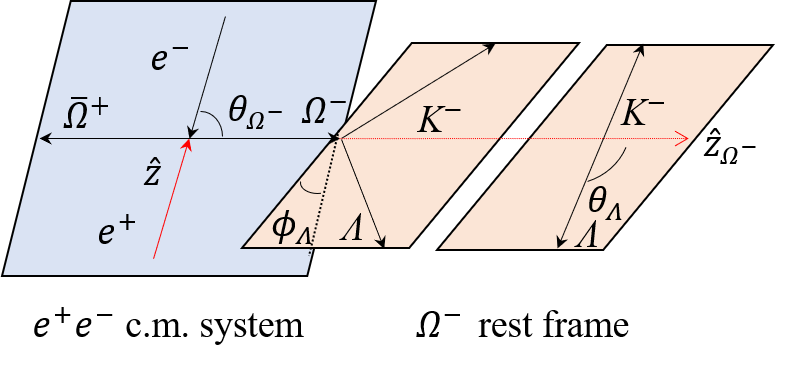}\\
   \vskip5pt
   \includegraphics[width=0.45\textwidth]{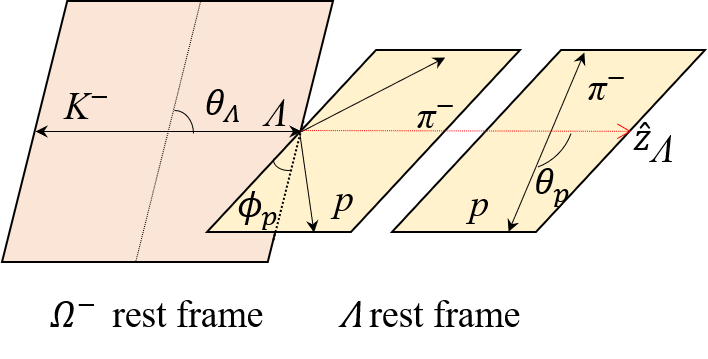}%
  \caption{Definition of the helicity angles used in the analysis. The helicity angles $\theta_{\Omega^{-}}$, $\theta_{\Lambda}$, $\phi_{\Lambda}$, $\theta_{p}$ and $\phi_{p}$ are spherical coordinates of the $\Omega^{-}$, $\Lambda$ and $p$ momenta in three reference frames: the  $e^+e^-$ c.m.\ system and the
  $\Omega^{-}$ and $\Lambda$ rest frames, respectively. The $\widehat{z}$-axis in the $e^+e^-$ c.m.\ system points along the incoming positron and $\widehat{z}_{\Omega^{-}}$ is the $\Omega^-$
  momentum direction. The polar axis direction in the $\Omega^-$ rest frame is $\widehat{z}_{\Omega^{-}}$ and
  $\widehat{y}_{\Omega^{-}}$ is along $\widehat{z}\times
  \widehat{z}_{\Omega^{-}}$, where $\widehat{z}_{\Lambda}$ is the
  $\Lambda$ momentum direction. The polar axis direction in the $\Lambda$ rest frame is $\widehat{z}_{\Lambda}$ and $\widehat{y}_\Lambda$ is along $\widehat{z}_{\Omega^{-}}\times \widehat{z}_{\Lambda}$. The decay of the $\bar{\Omega}^{+}$ follows the same principle.} \label{helicity_angle}
\end{figure}

In this Letter, using $(2.71 \pm 0.01) \times 10^9$ $\psip$ events from the BESIII detector~\cite{BESIII:2009fln,BESIII:2024lks}, we derive the joint angular distribution for the spin-3/2 $\Omega^{-}\bar{\Omega}^{+}$ entanglement from $\psip$ decays. Reconstructing the decays $\Omega^{-} \to K^{-}\Lambda (\to p\pi^{-})$ and $\bar{\Omega}^{+} \to K^{+}\bar{\Lambda} (\to \bar{p}\pi^{+})$ yields about 3300 events, enabling the first independent measurement of $\phi_{\Omega^{-}}$ and $\phi_{\bar\Omega^{+}}$. Using the joint angular distribution of $\xi = (\theta_{\Omega^{-}}, \theta_\Lambda, \phi_\Lambda, \theta_p, \phi_p,\theta_{\bar{\Lambda}}, \phi_{\bar{\Lambda}}, \theta_{\bar{p}}, \phi_{\bar{p}})$ (see Fig.~\ref{helicity_angle})~\cite{Perotti:2018wxm,Zhang:2023box,BESIII:2020lkm}, we probe CP symmetry and the weak and strong phase differences in $\Omega^{-}\to K^{-}\Lambda$ and $\bar{\Omega}^{+}\to K^{+}\bar{\Lambda}$ decay, and present the $\Omega^{-}\bar{\Omega}^{+}$ polarization correlations. The joint angular distribution is:
\begin{equation}\label{formulafinal}
\begin{aligned}
   \rho(\xi; H)&=\sum_{\mu,\bar{\mu}=0}^{15}\sum_{\nu,\bar{\nu}=0}^{3}C_{\mu,\bar{\mu}}b^{\Omega^{-}}_{\mu,\nu}b^{\bar{\Omega}^{+}}_{\bar{\mu},\bar{\nu}}a^{\Lambda}_{\nu,0}a^{\bar{\Lambda}}_{\bar{\nu},0},
\end{aligned}
\end{equation}
here $C_{\mu,\bar{\mu}}$, $b^{\Omega^{-}}_{\mu,\nu}$ , $b^{\bar{\Omega}^{+}}_{\bar{\mu},\bar{\nu}}$, $a^{\Lambda}_{\nu,0}$, and $a^{\bar{\Lambda}}_{\bar{\nu},0}$ are expressed in terms of the parameters: $H = (h_{i},\phi_{i} (i=1,3,4), \alpha_{\Omega^{-}}, \phi_{\Omega^-}, \alpha_{\bar{\Omega}^{+}}, \phi_{\bar{\Omega}^{+}}, \alpha_{\Lambda}, \alpha_{\bar{\Lambda}})$, as defined in Ref.~\cite{Zhang:2023box}. By fitting the joint angular distribution of the selected events with  Eq.~(\ref{formulafinal}), we can extract the helicity amplitudes and  $\Omega^{-}(\bar{\Omega}^{+})$ decay parameters. 

To ensure maximal signal purity, a double-tag method is employed by reconstructing both $\Omega^-$ and $\bar{\Omega}^+$ via their respective decay chains.  To improve the detection efficiency, we allow also one $\pi^{+}$ or $\pi^{-}$ to be missing per event.
All charged tracks reconstructed from multilayer drift chamber hits are required to be within a polar-angle ($\theta$) range of $|\cos\theta| < 0.93$. 
We require between five and seven such tracks. 
To determine the species of final-state particles, specific energy loss (d$E$/d$x$) and time-of-flight system (TOF) information are used to form particle identification (PID) probabilities for the reconstructed pion, kaon, and proton hypotheses. Charged particles are identified as the hypothesis with the highest probability, and only one $K^{-} (K^{+})$ and one proton (antiproton) are
required in each event. The remaining charged tracks in an event are assumed to be pions. 
Vertex fits are performed by looping over all combinations with oppositely charged proton and pion candidates, constraining them to a common vertex. If there is more than one $p\pi^-$ or $\bar{p}\pi^+$ pair, the pair with an invariant mass closest to the nominal $\Lambda$ mass~\cite{pdg} is selected. 
Given the possibility of an undetected  $\pi^{+}$ or $\pi^{-}$ in the decay,
events containing at least one surviving $\Lambda$ or $\bar{\Lambda}$ candidate are retained. 
In cases where both the $\pi^+$ and $\pi^-$ are found, one will be dropped and treated as missing, as discussed next, in order to obtain uniform mass resolution.  
To further suppress background and improve the kinematic resolution, a one-constraint (1C) kinematic fit is performed under the $p\bar{p}K^+K^-\pi\pi_{\rm miss}$ hypothesis, treating one pion as a missing particle. The mass of this $\pi_{\rm miss}$ is constrained to its Particle Data Group (PDG) value~\cite{pdg}.
For events with both $\Lambda$ and $\bar{\Lambda}$ fully reconstructed, the pion assigned as ``missing'' is chosen from the pair forming the $\Lambda$ or $\bar{\Lambda}$ candidate that has the smallest $\chi^{2}$ from its vertex fit.  Candidate events are required to satisfy $\chi^2_{1\rm C}$ $<$ 100. 
 In the following,
the $\pi_{\rm miss}$ momentum is that obtained from the 1C kinematic fit and is used in invariant-mass calculations.

The selection criteria require the invariant masses $M_{p\pi}$, $M_{p\pi_{\rm miss}}$, $M_{K p\pi}$, and $M_{K p\pi_{\rm miss}}$ to lie within [1.110, 1.122] GeV/$c^{2}$, [1.100, 1.132] GeV/$c^{2}$, [1.663, 1.681] GeV/$c^{2}$, and [1.655, 1.690] GeV/$c^{2}$, respectively. 
Finally, the candidates for the angular distribution analysis are obtained by constraining the invariant masses of $p\pi^{-}, \bar{p}\pi^{+}$ and $p\pi^{-}K^{-}, \bar{p}\pi^{+}K^{+}$ to the PDG values of the $\Lambda$ and $\Omega$ particles, respectively.

An inclusive $\psi(3686)$ Monte Carlo (MC) sample with  $2.3 \times 10^9$ $\psip$ events is used to study the possible background sources~\cite{Zhou:2020ksj} included in the simulation, and no peaking background is found.

\begin{figure}
     \begin{center}
    \includegraphics[width=0.45\textwidth]{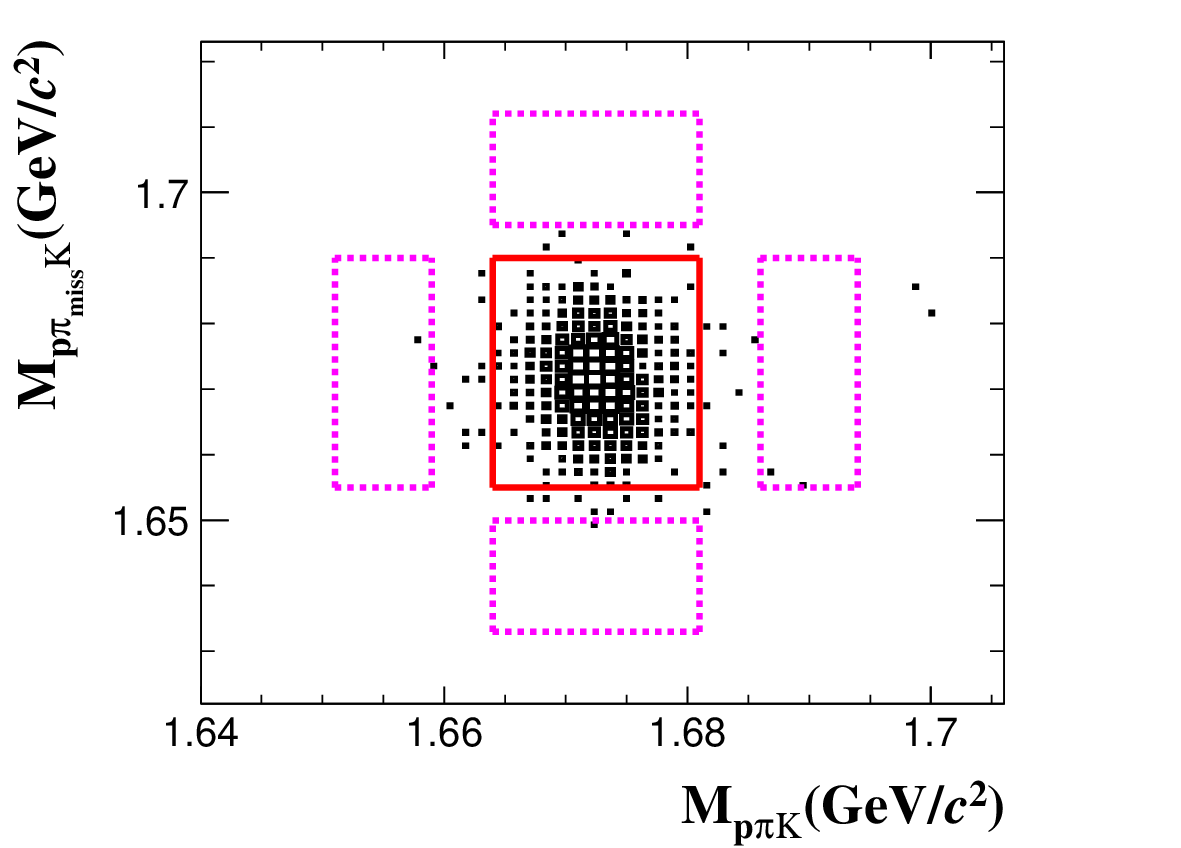}%
    \caption{Two-dimensional distribution of $M_{K p\pi}$ versus $M_{K p\pi_{\rm miss}}$ for selected $\psip \to \OOb$ candidates. The signal region is defined by the red solid box, while the sideband regions used for background estimation are outlined by the pink dashed boxes.}\label{sideband}
    \end{center}
    \end{figure}

Figure \ref{sideband} shows two-dimensional distribution of $M_{K p\pi}$ versus $M_{K p\pi_{\rm miss}}$ for the selected $K\Lambda$ candidates, revealing a clear cluster of $\psi(3686)\rightarrow\Omega^-\bar{\Omega}^{+}$ events within the red-boxed signal region. After applying all selection criteria, a total of 3272 events of $\psi(3686)\rightarrow\Omega^{-}\bar{\Omega}^{+}$ are obtained from the signal region for the angular distribution analysis. The non-$\Omega^{-}\bar{\Omega}^{+}$ background is estimated from predefined sideband regions in the $\Omega_{K p\pi}$ versus $\Omega_{K p\pi_{\rm miss}}$ plane (marked by pink dotted lines). A total of four background events are estimated from the sideband region, and scaled to the signal region by a factor equal to the ratio of the signal area to that of the four pink dashed sidebands.

 \begin{figure*}
    \begin{center}
    \includegraphics[width=0.24\textwidth]
    {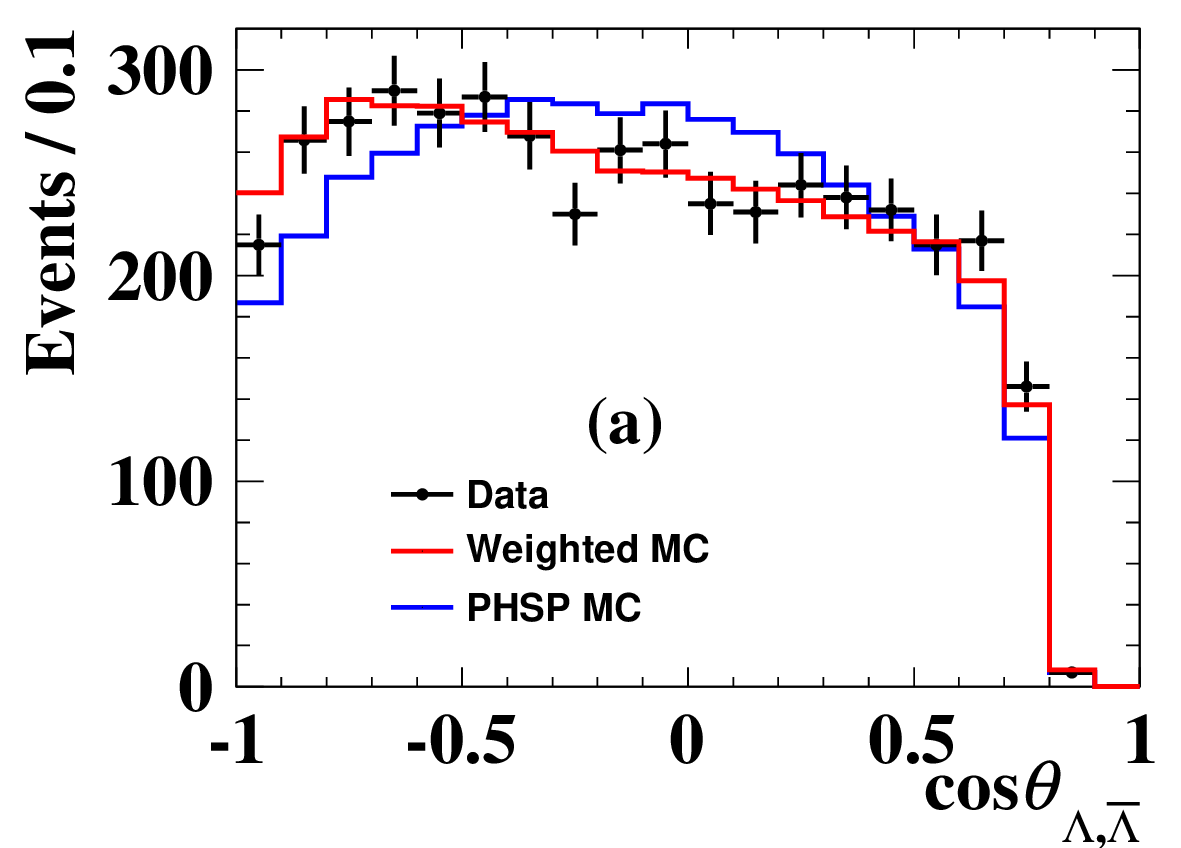}
    \includegraphics[width=0.24\textwidth]{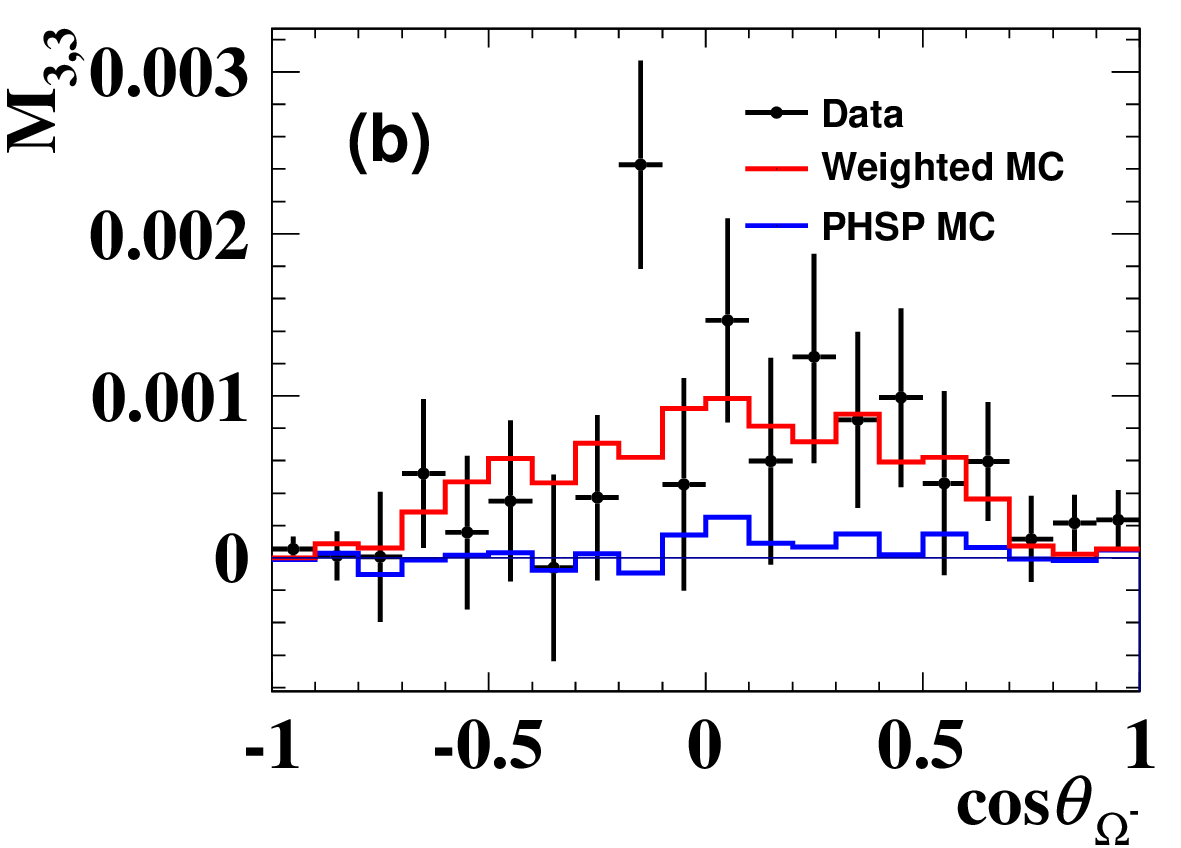}
    \includegraphics[width=0.24\textwidth]{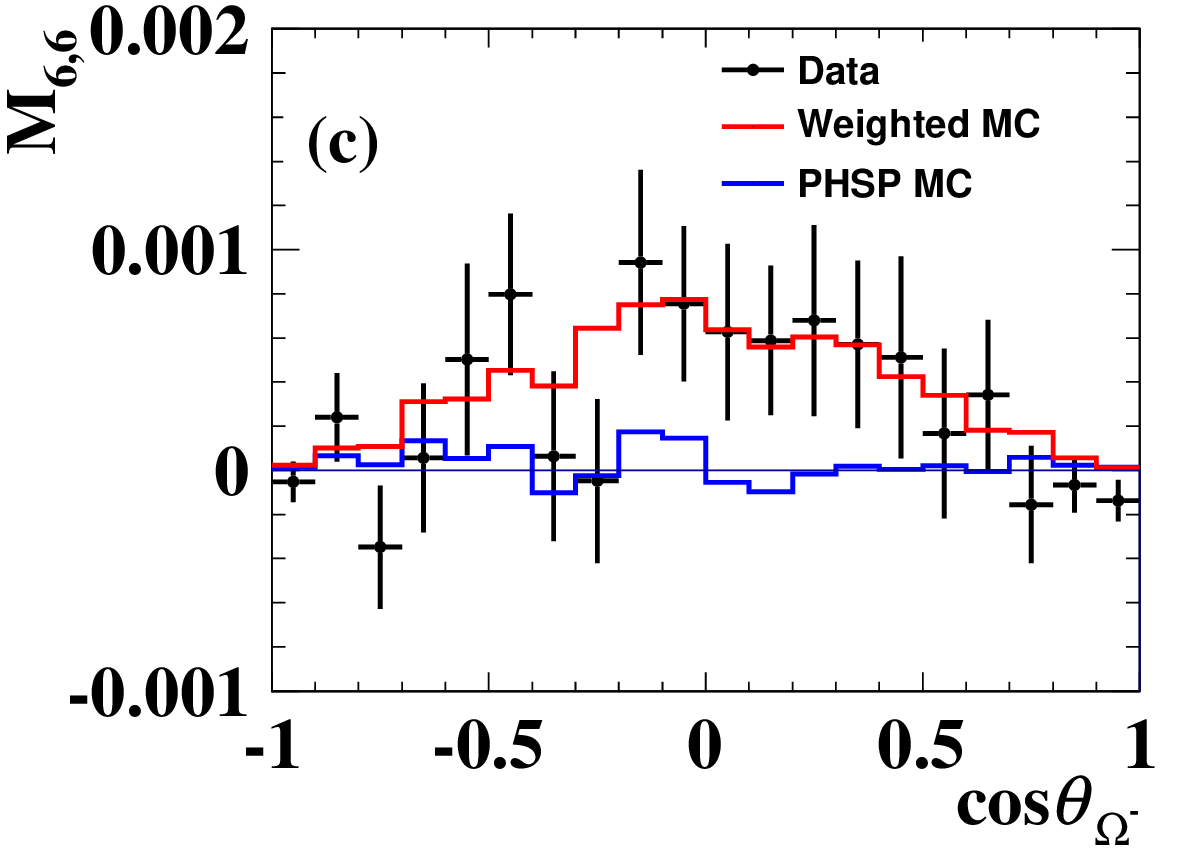}
    \includegraphics[width=0.255\textwidth]{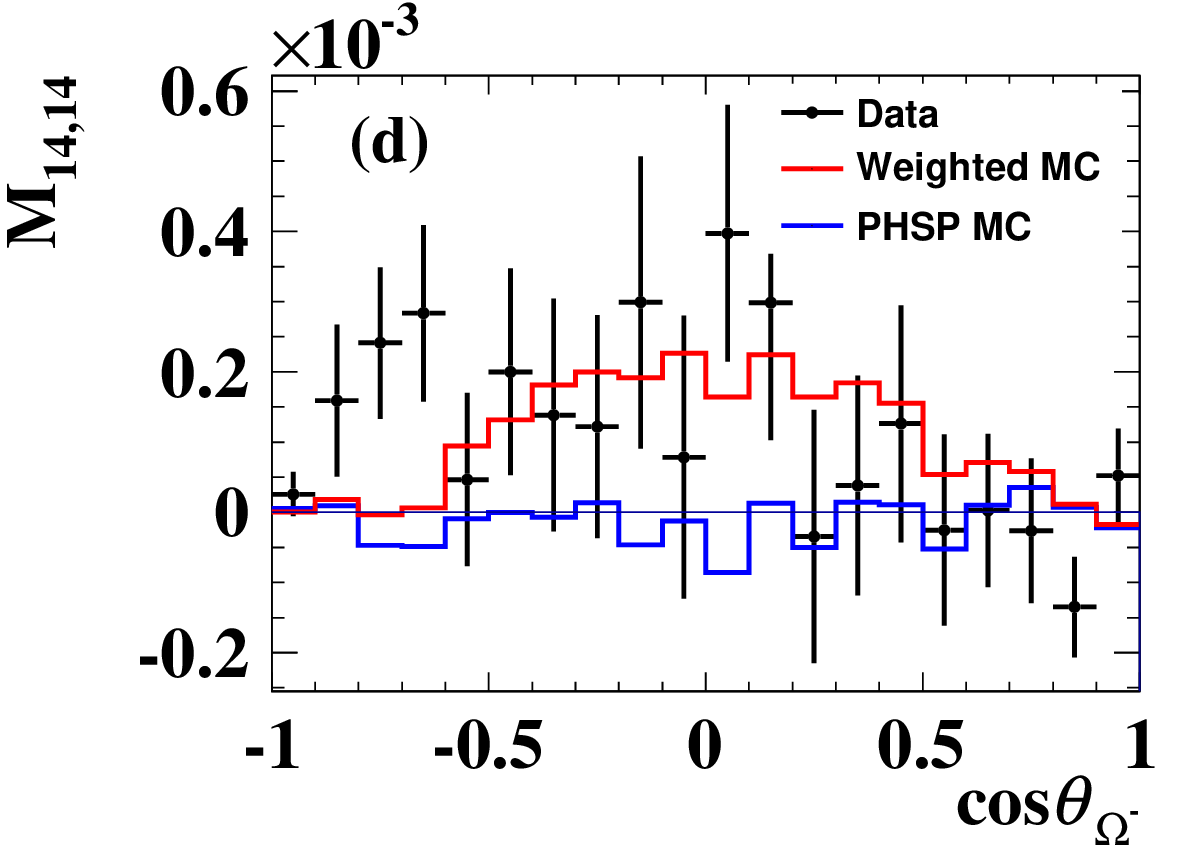}
    \hspace{1.5in}\parbox{6in}{\caption{(a) The $\cos\theta_{\Lambda,\bar{\Lambda}}$ distributions of data (dots with error bars), fits with the weighted PHSP MC sample (red histogram), and the PHSP MC sample (blue histogram); (b), (c), and (d) are the $M_{3,3}$, $M_{6,6}$, and $M_{14,14}$ distributions of data and fit results.}\label{moment_double}}
    \end{center}
\end{figure*}

To extract the decay parameters, an unbinned maximum-likelihood fit on the joint angular distribution is performed to the selected events.
The likelihood function is defined as:
\begin{equation}\label{like_function}
\begin{aligned}
\mathcal{L} = \prod_{j=1}^{N_{\rm sig}} \frac{\rho(\xi_j; H) \, \varepsilon(\xi_j)}{N(H)},
\end{aligned}
\end{equation}
where $\rho(\xi_j; H)$ is the angular distribution function for the decay described by Eq.~(\ref{formulafinal}), and $\xi_j$ represents the angular distribution variables of event $j$. $H = (h_{i}, \phi_{i}, \alpha_{\Omega^{-}}, \phi_{\Omega^-}, \alpha_{\bar{\Omega}^{+}}, \phi_{\bar{\Omega}^{+}}, \alpha_{\Lambda}, \alpha_{\bar{\Lambda}})$
are the parameters to be determined from the fit, and $\varepsilon(\xi_j)$ is the detection efficiency. $N_{\rm sig}$ is the number of selected events in the signal region, and $N(H)$ is the normalization factor calculated using the MC integration method. Background contributions are accounted for using events in the $\Omega_{K p\pi}$ vs.~$\Omega_{K p\pi_{\rm miss}}$ sideband regions. The fit is performed by minimizing the objective function \(S = -(\ln\mathcal{L}_{\rm data} - \ln\mathcal{L}_{\rm bg})\), where \(\ln\mathcal{L}_{\rm data}\) is the log-likelihood for events selected in the signal region of the data sample, and \(\ln\mathcal{L}_{\rm bg}\) is the log-likelihood for background events estimated by the sideband method.

In the absence of CP violation in $\Lambda$ decays, we set $\alpha_\Lambda = -\alpha_{\bar{\Lambda}} = 0.755 \pm 0.004$, which is world average from the previous experimental results~\cite{pdg}. By minimizing $S$, the remaining 10 unknown parameters in $H$ are obtained. The fit results of helicity amplitudes and $\Omega^{-}(\bar{\Omega}^{+})$ decay parameters are summarized in Table~\ref{fitting_result}.

\begin{table}[!htb]
    \footnotesize
    \centering
    \renewcommand{\tablename}{Table}
    \caption{Fit results for the helicity parameters in $\psip\to\Omega^{-}\bar{\Omega}^{+}$. The first (second) uncertainties are statistical (systematic) and the  $\phi_i $ parameters are given in radians.}\label{fitting_result}
    \begin{tabular}{c c}
    \hline
    \hline
    Parameter & Fit result \\
    \hline
     $h_{1}$         & $\phantom{-}0.60\pm0.05\pm0.03$\\
    $\phi_{1}$       & $\phantom{-}0.86\pm0.11\pm0.05$\\
    $h_{3}$          & $\phantom{-}0.49\pm0.05\pm0.02$\\
    $\phi_{3}$       & $\phantom{-}2.18\pm0.07\pm0.04$\\
    $h_{4}$          & $\phantom{-}0.66\pm0.03\pm0.01$\\
    $\phi_{4}$       & $\phantom{-}3.27\pm0.11\pm0.05$\\
    $\phi_{\Omega^{-}}$         & $-0.03\pm0.08\pm0.01$\\
    $\phi_{\bar{\Omega}^{+}}$   & $\phantom{-}0.03\pm0.08\pm0.03$\\
     $\alpha_{\Omega^{-}}$      & $-0.02\pm0.03\pm0.02$\\
    $\alpha_{\bar{\Omega}^{+}}$ & $-0.06\pm0.03\pm0.02$\\
    \hline
    \hline
    \end{tabular}
\end{table}

Signal MC events are generated according to the phase space (PHSP) distribution and then weighted using the matrix elements derived from the fit parameters. The predictions from this weighted MC sample are subsequently compared with data across five helicity angle distributions, incorporating background contributions estimated from the two-dimensional sidebands of $\Omega^{-}$ and $\bar{\Omega}^{+}$. We observe that the weighted PHSP describes data very well, while the PHSP sample fails to describe data, as shown in Fig.~\ref{moment_double}(a) for $\cos{\theta_{\Lambda,\bar{\Lambda}}}$, which has the most prominent difference. The moments $M_{3,3}$, $M_{6,6}$, and $M_{14,14}$ defined as $M_{\mu\bar{\mu}}=\frac{1}{N}\sum_{j=1}^{N}\sum_{\nu=0}^{3}\sum_{\bar{\nu}=0}^{3}b_{\mu,\nu}b_{\bar{\mu},\bar{\nu}}a_{\nu,0}a_{\bar{\nu},0}$ are compared between data and weighted MC samples, as shown in Figs.~\ref{moment_double}(b,c,d). Here $N$ is the number of events in the data samples. A clear preference for the weighted PHSP sample over the PHSP sample is 
observed.

The following systematic uncertainties are considered for the angular distribution measurement. The tracking and PID efficiencies for $K^{\pm}$ and the reconstruction efficiencies for $\Lambda, \bar{\Lambda}$ are studied using the control samples of $J/\psi \to p K^{-}\bar{\Lambda} + \text{c.c.}$ and $\Lambda^{+}_{c}\to\Lambda X$\cite{BESIII:2026mye}. The efficiency dependencies are parameterized in two-dimensional space: as a function of ($\cos\theta$, $p_{\text{t}}$) for kaons, and ($\cos\theta$, $p$) for $\Lambda, \bar{\Lambda}$. The MC efficiency is then corrected by applying these two-dimensional efficiency scale factors. The associated systematic uncertainty is estimated by varying the scale factors by $\pm 1\sigma$ independently within each corresponding two-dimensional bin. The differences between the new fit results and the nominal results are treated as systematic uncertainties. Systematic uncertainties from fixed parameters $\alpha_{\Lambda}$ and $\alpha_{\bar{\Lambda}}$ were estimated by varying their values by $\pm 1\sigma$ and taking fitting result differences as uncertainties. The systematic uncertainty of the fit method is estimated from input-output checks using 200 MC samples, each with a size equivalent to the data sample, generated with the nominal fit parameters from the angular analysis~\cite{Ping:2008zz}. The uncertainty is taken as the mean of the pull distribution scaled by the statistical uncertainty. The systematic uncertainties arising from the kinematic fits are examined by comparing the fitted decay parameters obtained with and without helix parameter corrections. All above contributions are added in quadrature to obtain the total systematic uncertainties as shown in Table~\ref{angular_distribution}.

\begin{table}[!htb]
   \scriptsize
   \centering
   \renewcommand{\tablename}{Table}
   \caption{Summary of the absolute systematic uncertainties for the decay parameters of $\psip\to\Omega^{-}\bar{\Omega}^{+}$. $\phi$ parameters are in radians.}\label{angular_distribution}
   \begin{tabular}{cccccc}
    \hline
    \hline
    Parameter              & Tracking, PID    & $\alpha_{\Lambda}, \alpha_{\bar{\Lambda}}$  &Fit method  &Kinematic & Total \\
    \hline
      $h_{1}$     &0.01             &0.01                  &0.03   &0.00        &0.03\\
     $\phi_{1}$    &0.04            &0.00                  & 0.02    &0.01     &0.05 \\
     $h_{3}$     &0.01             &0.00                  & 0.02      &0.00     &0.02\\
     $\phi_{3}$    &0.03             &0.00                  & 0.03       &0.01     &0.04 \\
     $h_{4}$     &0.01             &0.00                  &0.01        &0.01       &0.01\\
     $\phi_{4}$    &0.04            &0.00                  & 0.01        &0.03       &0.05 \\
     $\phi_{\Omega^{-}}$ &0.01           &0.00                  &0.00       &0.00          &0.01\\
     $\phi_{\bar{\Omega}^{+}}$ &0.03          &0.00                  & 0.00        &0.00   &0.03\\
    $\alpha_{\Omega^{-}}$ &0.02           &0.00                  &0.00              &0.01        &0.02\\
     $\alpha_{\bar{\Omega}^{+}}$ &0.02           &0.00                  &0.00     &0.00      &0.02\\
  \hline  
   \hline
  \end{tabular}
\end{table}

From Table~\ref{fitting_result}, we report the first measurements of the CP-sensitive observable $\phi_{\rm CP} = (-0.004\pm0.055\pm0.017)~\text{rad}$, and the phase differences $(\xi_{D}-\xi_{P}) = (-0.15 \pm 2.25 \pm 0.69)~\text{rad}$ and $(\delta_D - \delta_P) = (-0.97 \pm 0.88 \pm 0.34)~\text{rad}$, via Eq.~(\ref{weakCP}). We also obtain \(A_{\rm CP} = -1.63\pm1.86\pm1.07\) by adopting the PDG values~\cite{pdg}, which is consistent with but less precise than previous results~\cite{Chen:2005aza,Lu:2005fc,Lu:2006bn}. All CP symmetry results are summarized in Table~\ref{CPV_result}.


\begin{table}[!htb]
   \footnotesize
   \centering
   \renewcommand{\tablename}{Table}
   \caption{The CP symmetries in $\Omega\to \Lambda K$ decays. The first (second) uncertainties are statistical (systematic).  The angles $\phi, \delta, \xi$ are given in radians. }\label{CPV_result}
   \begin{tabular}{ccc}
   \hline
   \hline
    CP symmetry  &  This work & Previous result\\
     \hline
     $\phi_{\rm CP}$   & $-$0.004$\pm$0.055$\pm$0.017 &-\\
     $A_{\rm CP}$   & $-$1.63$\pm$1.86$\pm$1.07 &$-$0.02$\pm$0.13~\cite{Lu:2006bn}\\
      $B_{\rm CP}$   & $-$0.15$\pm$2.30$\pm$0.70 &-\\
      $\delta_{D}-\delta_{P}$ &  $-0.97\pm0.88\pm0.34$ &-\\
       $\xi_{D}-\xi_{P}$ &  $-0.15\pm2.25\pm0.69$ &-\\
      \hline
      \hline
  \end{tabular}
\end{table}

The value of $\phi_{\Omega^{-}}$ provides information on whether the process is $P$-wave ($\phi_{\Omega^{-}} = 0$) or $D$-wave ($\phi_{\Omega^{-}} = \pi$) dominant. By comparing the likelihood values obtained from fits fixing 
$\phi_{\Omega^{-}}=0$ or $\pi$ against the nominal fit, we find that the P-wave hypothesis is disfavored at the 0.3$\sigma$ level, whereas the D-wave hypothesis is disfavored at the 22$\sigma$ level. This favors the $P$-wave dominant scenario, consistent with theoretical predictions~\cite{Tandean:2004mv}. 


\begin{figure*}
\centering
\includegraphics[width=0.32\textwidth]{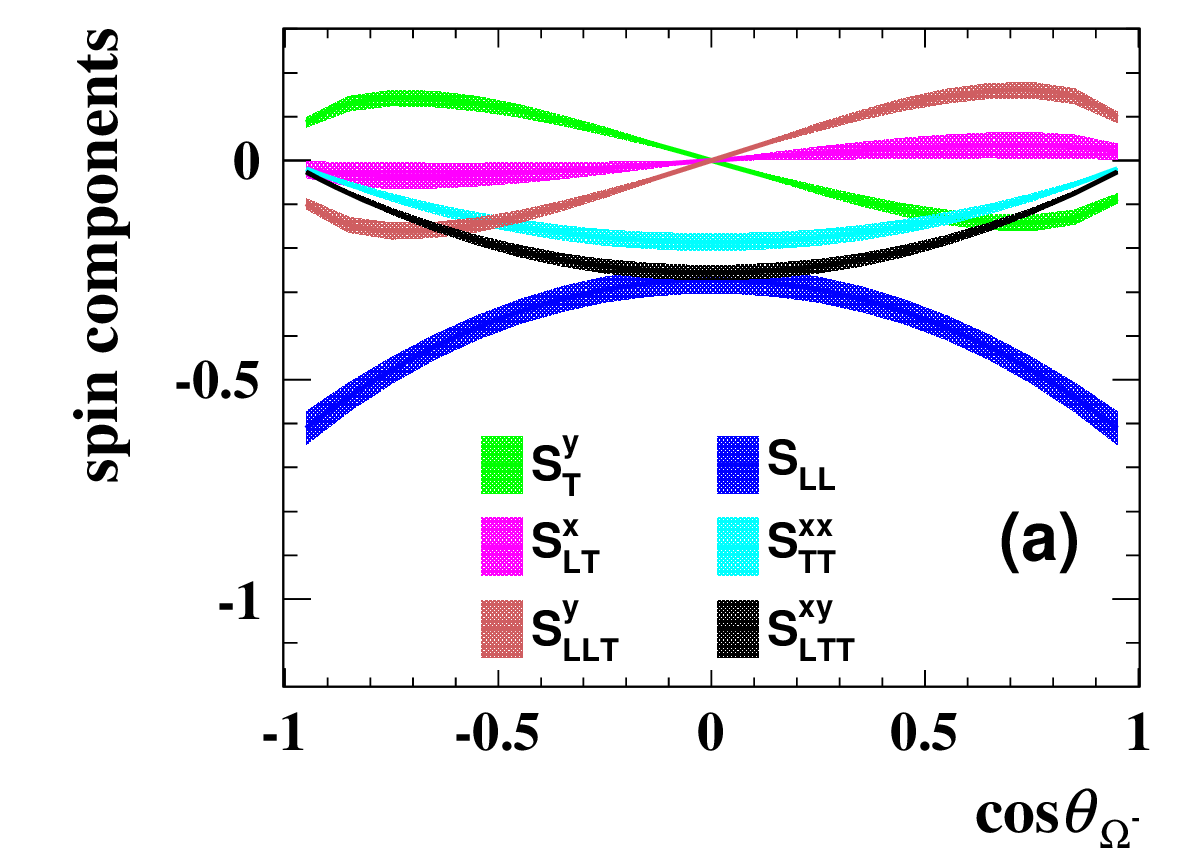}
\includegraphics[width=0.32\textwidth]{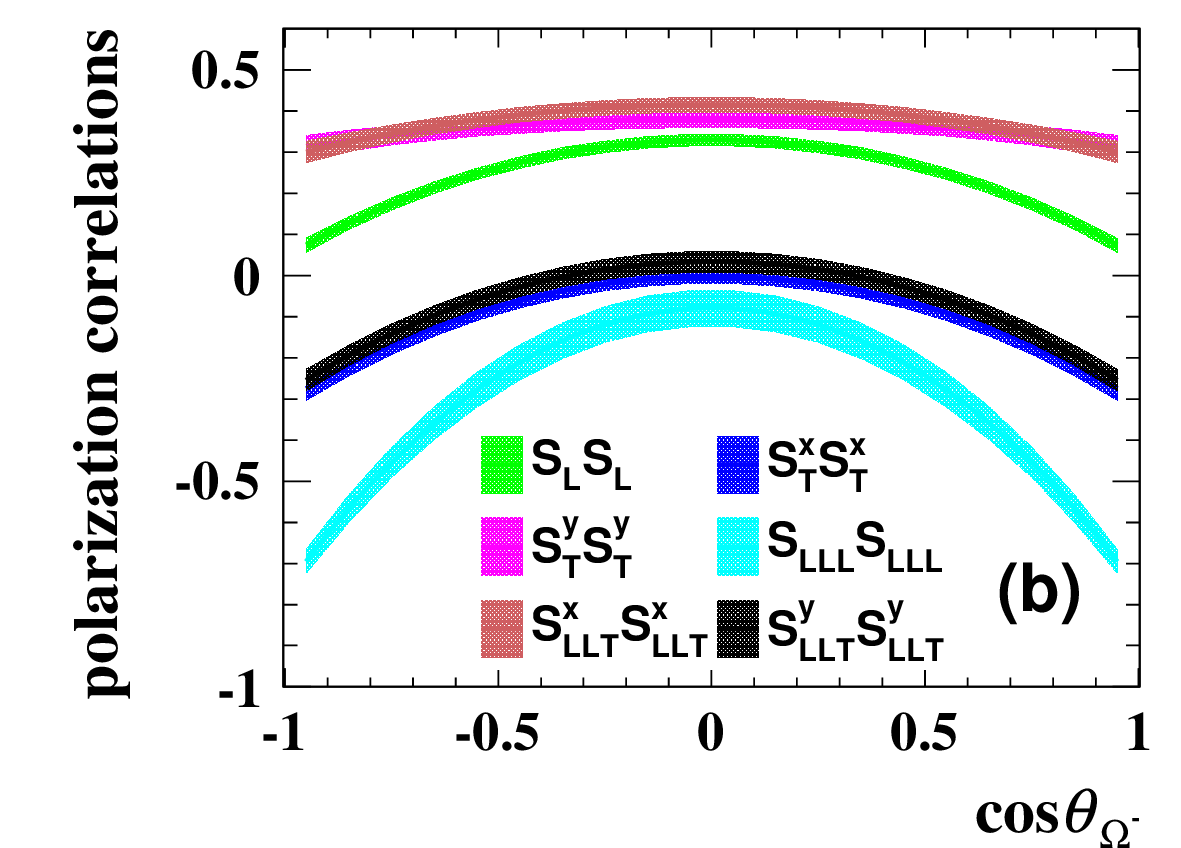}
\includegraphics[width=0.32\textwidth]{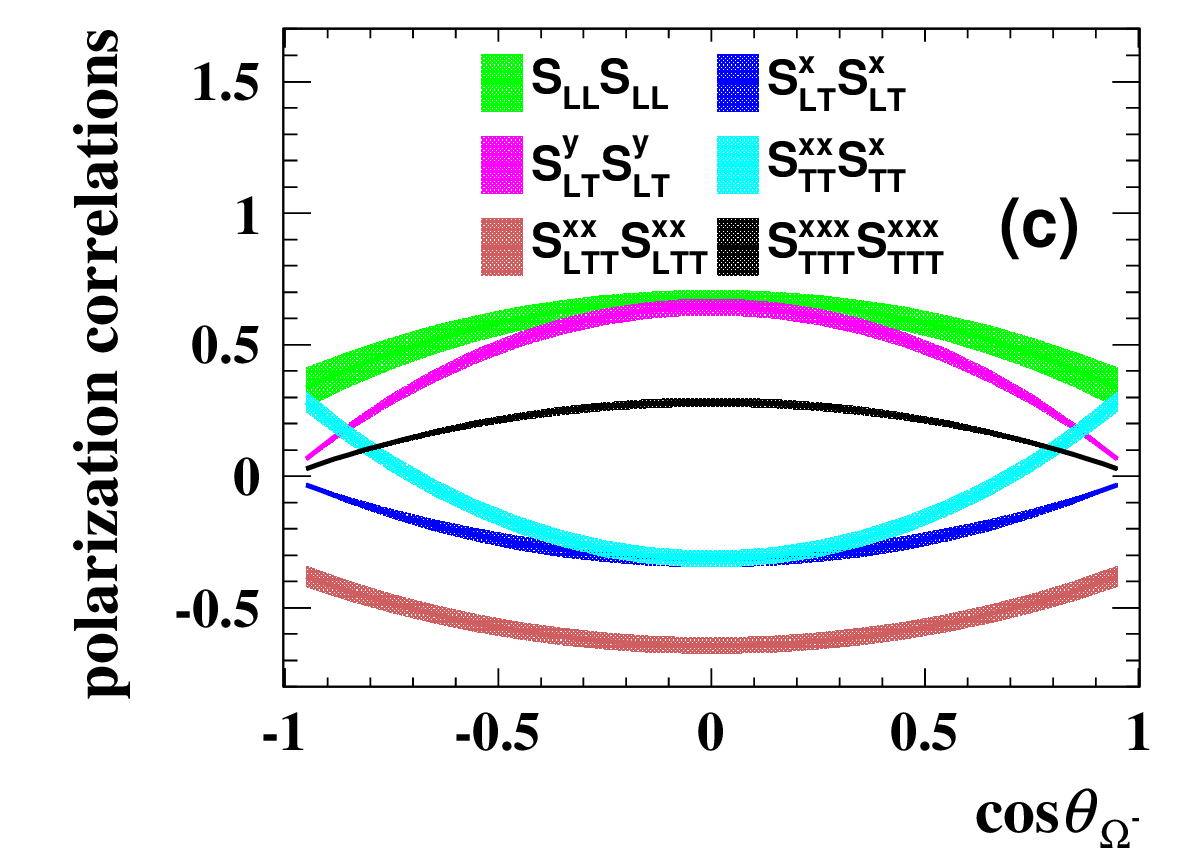}
\caption{(a) The $\cos\theta_{\Omega}$ dependence of the multi-polar polarization components; (b), (c) some polarization correlations between $\Omega^{-}$ and $\bar{\Omega}^{+}$. The solid lines represent the central values, and the shaded areas represent $\pm 1\sigma$ as calculated from the covariance matrix of the fitted $h_{i}$ and $\phi_{i}$.}
\label{polarization}
\end{figure*}

Based on the helicity amplitudes from Table~\ref{fitting_result}, we present the $\cos\theta_{\Omega}$-dependence of the single $\Omega^{-}$ polarization in Fig.~\ref{polarization}(a), which exhibits significant vector ($S_T^y$) and tensor (rank-2: $S_{LL}$, $S_{TT}^{xx}$; rank-3: $S_{LLT}^y$, $S_{LTT}^{xy}$) components~\cite{Zhang:2023box,Zhang:2023wmd}. In contrast, exploiting the entangled $\Omega^-\bar{\Omega}^+$ pair, we show the corresponding polarization correlations in Figs.~\ref{polarization}(b) and \ref{polarization}(c). As seen in Fig.~\ref{polarization}(b), several correlations vary strongly with $\cos\theta_{\Omega}$, with sign alternations indicative of varying constructive and destructive interference. Notably, some longitudinal tensor correlations show particularly clear angular variations. Figure~\ref{polarization}(c) further shows angular variations in other tensor correlations, presenting either positive or negative values~\cite{Zhang:2023box}. In all panels, the solid curves denote the central values, and the shaded bands represent $\pm 1\sigma$ uncertainties, propagated from the covariance matrix of the fitted parameters $h_i$ and $\phi_i$.

Overall, this analysis provides a foundation for future hyperon CPV searches by demonstrating the power of full angular distribution measurements in $\Omega^{-}\bar{\Omega}^{+}$ decays. We reconstruct spin-entangled $\Omega^{-}\bar{\Omega}^{+}$ pairs to extract the complete angular distribution, yielding the CP-sensitive observables $\phi_{\Omega^{-}}$ and $\phi_{\bar{\Omega}^{+}}$, as well as the full set of polarization correlations (including vector, tensor, and octupole components) between the two hyperons. This approach enables improved CPV measurements and provides access to $\Omega^{-}$ polarization properties and their entanglement with $\bar{\Omega}^{+}$. Although the $\Omega^{-}\to \Xi\pi$ channel accounts for approximately 30\% of $\Omega^{-}$ decays and could offer complementary CPV signatures due to its distinct strong-phase behavior, we focus exclusively on the $\Omega^{-}\to K^{-}\Lambda$ channel in this initial analysis due to favorable statistics. Future studies incorporating multi-channel interference effects are expected to further enhance sensitivity. Notably, excluding the previously measured \(A_{\rm CP}\), all CP asymmetry observables and the \(\Omega^-\bar{\Omega}^+\) polarization correlations presented in this work are measured for the first time.

\begin{center}ACKNOWLEDGMENTS\end{center}

The BESIII Collaboration thanks the staff of BEPCII (https://cstr.cn/31109.02.BEPC) and the IHEP computing center for their strong support. This work is supported in part by National Key R\&D Program of China under Contracts Nos.2023YFA1609400, 2023YFA1606000, 2023YFA1606704, 2025YFA1613900; National Natural Science Foundation of China (NSFC) under Contracts Nos. 12305085, 11635010, 11935015, 11935016, 11935018, 12025502, 12035009, 12035013, 12061131003, 12192260, 12192261, 12192262, 12192263, 12192264, 12192265, 12221005, 12225509, 12235017, 12342502, 12361141819, 12535005, 12205255; the Natural Science Foundation of Henan Province No. 252300421214; the Chinese Academy of Sciences (CAS) Large-Scale Scientific Facility Program; the Strategic Priority Research Program of Chinese Academy of Sciences under Contract No. XDA0480600; CAS under Contract No. YSBR-101; 100 Talents Program of CAS; The Institute of Nuclear and Particle Physics (INPAC) and Shanghai Key Laboratory for Particle Physics and Cosmology; Agencia Nacional de Investigaci\'on y Desarrollo de Chile (ANID), Chile under Contract No. ANID CCTVal CIA250027; ERC under Contract No. 758462; German Research Foundation DFG under Contract No. FOR5327; Istituto Nazionale di Fisica Nucleare, Italy; Knut and Alice Wallenberg Foundation under Contracts Nos. 2021.0174, 2021.0299, 2023.0315; Ministry of Development of Turkey under Contract No. DPT2006K-120470; National Research Foundation of Korea under Contract No. RS-2026-25486791; National Science and Technology fund of Mongolia; Polish National Science Centre under Contract No. 2024/53/B/ST2/00975; STFC (United Kingdom); Swedish Research Council under Contract No. 2019.04595; U. S. Department of Energy under Contract No. DE-FG02-05ER41374




\end{document}